\documentclass[twocolumn,showpacs,preprintnumbers,amsmath,amssymb]{revtex4}

\usepackage{graphicx}
\usepackage{dcolumn}

\begin{document}

\title{Structure of a large social network}

\author{G\'abor Cs\'anyi\footnote{Corresponding author}}
\email{gabor@csanyi.net}
\affiliation{TCM Group, Cavendish Laboratory, University of Cambridge \\
Madingley Road, Cambridge CB3 0HE, United Kingdom}

\author{Bal\'azs Szendr\H oi}
\email{szendroi@math.uu.nl}
\affiliation{Department of Mathematics, Utrecht University \\ PO. Box 80010,
NL-3508 TA Utrecht, The Netherlands}

\date{\today}

\begin{abstract}
We study a social network consisting of over $10^4$ individuals, with
a degree distribution exhibiting two power scaling regimes separated
by a critical degree $k_{\rm crit}$, and a power law relation between
degree and local clustering. We introduce a growing random model based
on a local interaction mechanism that reproduces all of the observed
scaling features and their exponents. Our results lend strong support
to the idea that several very different networks are simultenously
present in the human social network, and these need to be taken into account
for successful modeling.

\end{abstract}

\maketitle

%
%

The ubiquity of networks has long been appreciated: complex
systems in the social and physical sciences can often be modelled on a
graph of nodes connected by edges. Recently it has also been realized
that many networks arising in nature and society, such as neural
networks~\cite{watts_strog}, food webs~\cite{food}, 
cellular networks~\cite{jeong2000}, networks of sexual
relationships~\cite{sex}, collaborations between film actors
\cite{watts_strog,ba} and scientists~\cite{new_pnas,barab2001},
power grids~\cite{watts_strog,classes}, Internet routers~\cite{faloutsos} 
and links between pages of the World Wide Web~\cite{ajb,kumar,broder} 
all share certain universal characteristics very poorly modelled by 
random graphs~\cite{er}: they are highly clustered 
``small worlds''~\cite{milgram, watts_strog, watts} with
small average path length between nodes, and they
have many highly connected nodes with a degree distribution often
following a power law~\cite{ajb, ba}. The network of humans with
links given by acquaintance ties is one of the most intriguing of such
networks~\cite{milgram, pool_kochen, was_faust, watts}, but its study has
been hindered by the absence of large reliable data sets. 

%
%

\noindent
{\it --- The new data set\ } The WIW project was started by a small group
of young professionals in Budapest, Hungary in April 2002 on the web
site www.wiw.hu with the aim to record social acquaintance. The
network is invitation--only and new members join by an initial link
connecting to the person who invited them. New links are recorded
between members after mutual agreement. This scheme results in members
preferring to use their real names and effectively prevents
proliferation of multiple pseudonyms.  Because of the relatively short
age of the network, links formed between people newly acquainted
through the web site have a minimal structural effect; thus the
majority of the links represent genuine social acquaintance and the
WIW develops as a growing subgraph of the underlying social
acquaintance network.  Indeed, the growth process of the WIW network
is essentially equivalent to the ``snowball sampling'' method well
known to sociologists~\cite{was_faust}, and to the crawling methods
used to investigate the World Wide Web and other computer networks.
We study the WIW network using two anonymous snapshots taken in
October 2002 (with 12388 nodes and 74495 links) and January 2003 (with
17496 nodes and 127190 links).

%
%

\begin{figure}
\scalebox{0.40}{\includegraphics{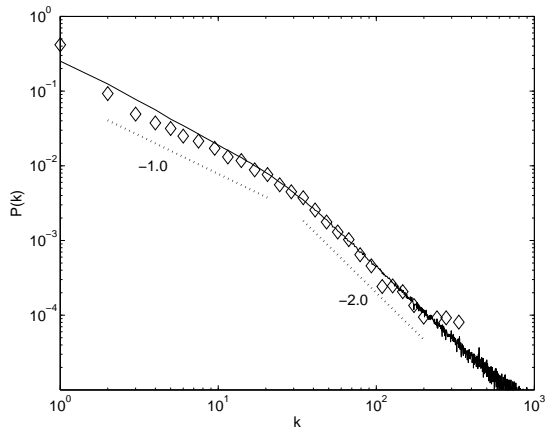}}
\caption{The degree distribution of the WIW network (diamonds), with a small-$k$
power law $P(k)\sim k^{-1.0}$ and a large-$k$ power law $P(k)\sim k^{-2.0}$
separated by a critical degree $k_{\rm crit}\approx 25$.
The solid line gives the degree distribution of the model of the text with 
edge/node ratio $m=15$, $q=0.5$ and size $V=2\times 10^5$, averaged over 
50 graphs.}
\label{deg}
\end{figure}

The degree distribution of the WIW network is plotted on
Figure~\ref{deg}.  The graph shows two power law regimes
\[P(k) \sim \left\{\begin{array}{cl} {k^{-1.0}}& \mbox{ if } k<k_{\rm crit} \\ {k^{-2.0}}& \mbox{ if } k>k_{\rm crit} \end{array}\right.\]
The two regimes are separated by a critical degree
$k_{\rm crit}\approx 25$. The exponent $\gamma_2\approx -2$ of the
large-$k$ power law falls in a range that has often been observed
before in a variety of contexts~\cite{sex, ajb, kumar, broder, ba, faloutsos,
new_pnas, barab2001, classes, jeong2000}. The value
$\gamma_1\approx-1$ of the small-$k$ power law exponent is much less
common, observed before only in some scientific collaboration
networks~\cite{new_pnas} and food webs~\cite{food}. The possibility of
a double power law was discussed in~\cite{barab2001, new_coll}, but
the WIW network is the first data set which conclusively demonstrates
the existence of double power law behaviour. The two snapshots give
essentially identical distributions. Since the network grew by about
50\% during this period, the described distribution can be regarded as
essentially stationary in time.

%
%

\begin{figure}
\scalebox{0.40}{\includegraphics{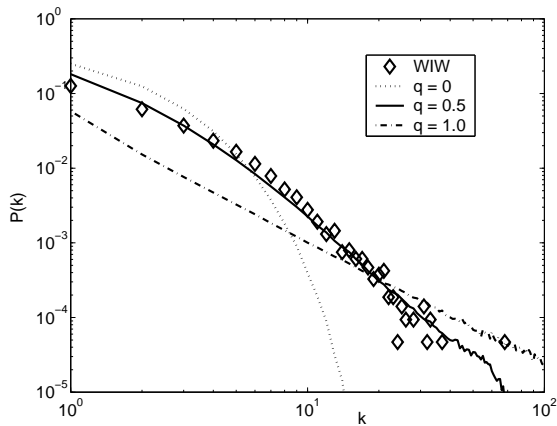}}
\caption{The degree distribution of the invitation tree of the WIW graph 
(diamonds) exhibits a large-$k$ power law $P(k)\sim k^{-3}$. Also plotted
is the invitation tree of the model graph with $V=2\cdot 10^5$ nodes, $m=15$ and
different parameters $q$.}
\label{invite_tree}
\end{figure}

The two scaling regimes in the degree distribution of the WIW graph are indicative 
of two distinct growth processes: the invitation of new members, and
the recording of acquaintance between already registered members. 
The degree distribution of the invitation tree graph is shown on
Figure~\ref{invite_tree}, where a power law is observed for
large degrees with an exponent $\gamma\approx-3$. Since this distribution 
is qualitatively different from the total degree distribution, it
is reasonable to conclude that there are at least two
different types of social linking at play here: the network of
friends defined by ties strong enough to warrant an
invitation is different from the network of
acquaintances that drives the mutual recognition, once both parties
are registered. 

%
%

\begin{figure}
\scalebox{0.40}{\includegraphics{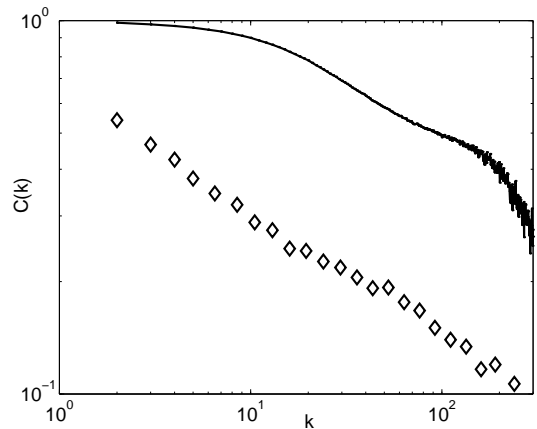}} 
\caption{The correlation between the local clustering coefficient $C(k)$ and
the node degree $k$ for the WIW graph (diamonds), showing a power
law $C(k)\sim k^{-0.33}$. The solid line plots the same for the model graph 
with parameters $V=2\cdot10^5$, $m=15$ and critical $q=0.5$, averaged over 
$50$ graphs.}
\label{kck}
\end{figure}

The density of edges in the neighbourhood of a node is measured 
by the local clustering coefficient. For a node $v$ of degree $k$,  
the local clustering coefficient $C(v)$ is the number of acquaintance
triangles of which $v$ is a vertex, divided by $k(k-1)/2$, the number of 
all possible triangles. Figure~\ref{kck} plots $C(k)$, 
the average of $C(v)$ over nodes of degree $k$, against the degree, showing
the existence of a power law \[C(k)\sim k^{-0.33}.\] 
A relationship $C(k)\sim k^{-\alpha}$
was observed before in~\cite{vazvesp, rav_barab}, 
but with significantly larger exponents. Such power laws hint at the presence 
of hierarchical architecture~\cite{rav_barab}: when small groups organize into 
increasingly larger groups in a hierarchical manner, the local clustering
decreases on different scales according to such a power law. 

The average clustering coefficient $\langle C\rangle\approx0.2$ is
obtained as the average of $C(v)$ over all nodes.  The average
diameter between two members of the WIW network is about $4.5$. These
two measures indicate the ``small--world'' nature of the WIW network in
the sense of~\cite{watts_strog}.

%
%

\vspace{0.1in}

\begin{figure}
\scalebox{0.4}{\includegraphics{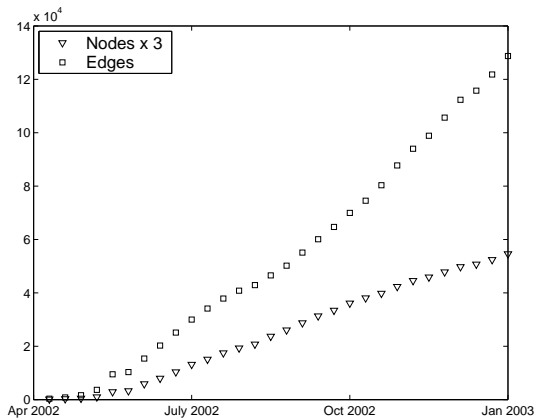}} 
\caption{The time development of the number of nodes and edges of the WIW network.
Note that the number of nodes is multiplied by $3$ for better visibility.}
\label{time}
\end{figure}

\noindent
{\it ---Time development\ } As Figure~\ref{time} shows, the number of
nodes of the WIW network grew approximately linearly with
time. This appears to be related to the fact that the WIW network
develops as a subgraph of the underlying social network, and thus the
availability of new members is constrained by high clustering of the
existing social links: a substantial proportion of the aquaintances of
a newly invited member will have been invited already.

On the other hand, the number of edges also grew linearly with time,
and thus the edge/node ratio only grew moderately during the
existence of the network. This observation is in contradiction with
any purely local time-independent edge creation mechanism. If every
member of the network actively participates in edge creation
indepently of its age in the network, the edge/node ratio would also
increase linearly with time. This linear growth of the edge/node
ratio was not observed in the network, and hence we conclude that the
edge creatioon activity of members necessarily decreased with time.

The fact that the edge/node ratio changes little over time is
consistent with the observed stationary nature of the degree
distribution. To see this, consider a growing network with $V(t)$
nodes at time $t$ and a time-independent degree distribution $P(k)$
with $\sum_k P(k)=1$ and finite first moment $\sum_k kP(k)$. At time $t$, 
there are \[n(k,t)=V(t)P(k)\] nodes of degree $k$, and
hence the number of edges is
\[E(t)=\frac{1}{2} \sum_k kn(k,t) = \frac{V(t)}{2}\sum_k kP(k).\]
Consequently the edge/node ratio $E(t)/V(t)$ is essentially constant, 
and it only changes because the maximal degree increases. This argument 
applies to the WIW network with stationary distribution
\[P(k) = \left\{\begin{array}{cl} \frac{c_1}{k^{\gamma_1}}& \mbox{ if } k<k_{\rm crit} \\ \frac{c_2}{k^{\gamma_2}}& \mbox{ if } k>k_{\rm crit} \end{array}\right.\]
with $\gamma_1\approx -1$, $\gamma_2\approx -2$. This distribution is
on the boundary of distributions with finite first moment: the first
moment exists for $\gamma_2<-2$ but not otherwise.

\vspace{0.1in}

\noindent {\it ---Network modelling\ }
A random graph process based on linear preferential attachment for the
creation of new edges was proposed in~\cite{ba} to account for the
observed power laws in natural networks. Such a process leads indeed
to a graph with a power law degree distribution~\cite{ba, bollob}. 
However, this model is by definition macroscopic, requiring
information about the entire network in every step.  This assumption is 
realistic for the World Wide Web or some collaboration networks,
where all nodes are ``visible'' from all others. For human
social networks however, it is reasonable to assume some degree of locality 
in the interactions.  Also, the original scale-free models are not
applicable to networks with high, degree dependent clustering
coefficients. These problems motivated the introduction of new models
which use local triangle creation mechanisms~\cite{watts, hk, vazquez, deb, jin},
which increase clustering in the network.  These models have
degree-dependent local clustering, and can also lead to
power law degree distributions, though no existing model of this kind
shows a double power law. 

We now present a new model to account for the observed properties of
the WIW network. As mentioned above, the WIW can be viewed as a
growing subgraph of the underlying social acquiantance graph. This
suggests a model obtained by a two-step process, first modelling the
underlying graph, and then implementing a growth process. The lack of
available data on the underlying graph however prevents us from
following this programme directly. We build instead a growing graph in one 
single process, choosing the local triangle mechanism as our basic edge creation 
method. This models the social introduction of two members 
of the WIW network by a common friend some time in the past, such edges being 
gradually recorded in the WIW network itself. The invitation of new members is 
modelled by sub-linear preferential attachment~\cite{ba}, motivated by experimental 
results on scientific collaboration networks~\cite{barab2001,new_timedev}, where
the data permits the analysis of initial edge formation. We also impose the
constant edge/node ratio to be consistent with the observed stationary 
distribution. Note that this constant has to be tuned 
from the shape of observed distributions, and cannot be inferred from the WIW data 
directly. The reason for this is that the WIW has a disproportionate number of 
nodes of degree one (Figure~\ref{deg}), representing people who once responded to 
the invitation but never returned, which distorts the edge/node ratio 
without invalidating our other conclusions.  

The precise description of our process is as follows. 
\begin{itemize}
\item We begin with a small regular graph.
\item New nodes arrive at a rate of one per unit time and
are attached to an earlier node chosen with a probability
distribution giving weight~$k^q$ to a node of degree~$k$, where
$q\geq 0$ is a parameter.
\item Internal edges are created as follows: we select two random 
neighbours of a randomly chosen node $v$, and if they are unconnected, 
we create an edge between them. Otherwise, we select two new neighbours
of the same node $v$ and try again. 
\item A constant number of internal edges is created per unit time, so that 
the edge/node ratio equals a constant $m$ after each time step.
\end{itemize}

\begin{figure}
\scalebox{0.40}{\includegraphics{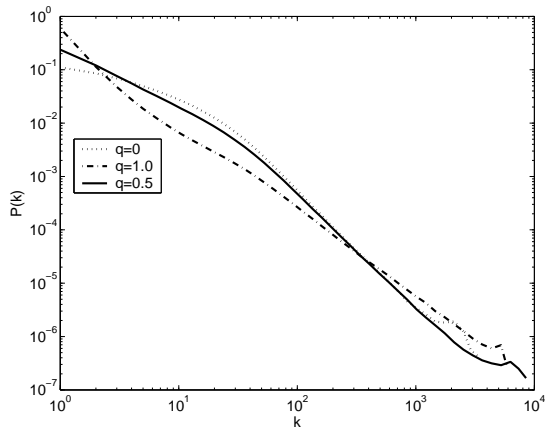}} 
\caption{The dependence of the degree distribution of our model graph on the
parameter $q$, with $m=15$ and $V=2\cdot 10^5$ ($q=0.5, 1$) and $V=5\cdot 10^4$ 
($q=0$), averaged over 50 graphs.}
\label{csz_q}
\end{figure}

The degree distribution of graphs generated by our process is shown on
Figure~\ref{csz_q}. We found a very robust large~$k$ power law of
exponent $\gamma_2\approx -2$, essentially independently of the
invitation mechanism.  We measured the joint probability distribution
of the degrees $k$,$k^\prime$ of nodes connected by new internal
edges, and found that for large values, it was proportional to
$kk^\prime$. This directly leads to a power law exponent of $-2$ via
standard mean-field arguments~\cite{barab2001}. The small~$k$
behaviour was found to be sensitive to the invitation mechanism;
Figure~\ref{csz_q} shows that a second power law only appears at a
critical $q$. The critial value of $q$ depends on the edge/node ratio.

To test the hypothesis that the low degree power law is indeed related
to the invitation mechanism, we plot on Figure~\ref{invite_tree} the
degree distribution of the invitation tree of the model network for
various values of the parameter $q$. At the critical value, we obtain
a scale-free distribution with exponent $-3$. Decreasing $q$ leads to
a much sharper drop in the curve, with an exponential tail for $q=0$,
whereas increasing $q$ above the critical value leads to a
gelation-type behaviour: new nodes connect only to very large degree
nodes.

Figure~\ref{deg} shows that, for appropriate choice of parameters, the degree 
distribution of our model reproduces that of the real network extremely well. 
Figure~\ref{kck} plots the dependence of the local clustering cofficient 
$C(k)$ as a function of the degree $k$ in our model network.  
While no simple power law can be observed, there is a clear trend of 
decreasing clustering with increasing degree, indicative of the presence 
of hierarchy in the model network~\cite{rav_barab}. 
 
\vspace{0.1in}

\noindent
{\it ---Conclusion\ } We have presented and analyzed a large new data set
of a human acquintance network with a stable degree distribution which
exhibits a new feature: two power law regimes with different
exponents. The observed approximately constant edge/node ratio is a
consequence of the stability of the degree distribution, and implies
that the average activity of members is time dependent, whereas
the growth of the number of nodes is constrained by  
the underlying social network.  We also introduced a
model which reproduces the observed degree distribution extremely well, and 
concluded that the small-$k$ power law is a related to the scale-free nature
of the invitation tree, whereas the large-$k$ power law is a result
of the triangle mechanism of social introductions. Our results show that 
human social networks are likely to be composed of several networks with different 
characteristics, and directly observable processes will exhibit a mixture
of features resulting from distinct underlying mechanisms. 

\begin{acknowledgments}
We thank Zsolt V\'arady and D\'aniel Varga for access to the data of the WIW
network, and Risi Kondor for helpful discussions. 
\end{acknowledgments}


\begin{thebibliography}{99}
\bibitem{watts_strog} D.J. Watts and S.H. Strogatz, Nature {\bf 393}, 440 (1998). 
\bibitem{food} J.M. Montoya and R.V. Sole, J. Theor. Biol. {\bf 214}, 405 (2002). 
\bibitem{jeong2000} H. Jeong {\it et al.}, Nature {\bf 407}, 651 (2000).
\bibitem{sex} F. Liljeros {\it et al.}, Nature {\bf 411}, 907 (2001).
\bibitem{ba} A.-L. Barab\'asi and R. Albert, Science {\bf 286}, 509 (1999).
\bibitem{new_pnas} M. E. J. Newman, {Proc. Natl. Acad. Sci. USA} {\bf 98}, 404 (2001).
\bibitem{barab2001} A.-L. Barab\'asi {\it et al.}, {Physica A} {\bf 311}, 590 (2002). 
\bibitem{classes} L.A.N. Amaral, A. Scala, M. Barthelemy, and H. E. Stanley, {Proc. Natl. Acad. Sci. USA} {\bf 97}, 11149 (2000).
\bibitem{faloutsos} M. Faloutsos, P. Faloutsos, and C. Faloutsos. Comp. Comm. Rev. {\bf 29}, 251 (1999).
\bibitem{ajb} R. Albert, H. Jeong, and A.-L. Barab\'asi, Nature {\bf 401}, 130 (1999). 
\bibitem{kumar} R. Kumar, P. Raghavan, S. Rajalopagan, and A. Tom\-kins, in {\it Proc. 9th ACM Symp. on Principles of Database Systems} (Association for Computing Machinery, New York, 2000) 1.
\bibitem{broder} A. Broder {\it et al.}, Comput. Netw. {\bf 33}, 309 (2000).
\bibitem{er} P. Erd\H os and A. R\'enyi, {Publ. Math. Debrecen} {\bf 6}, 290 (1959).
\bibitem{milgram} S. Milgram, {Phychol. Today} {\bf 2}, 60 (1967). 
\bibitem{watts} D. J. Watts, {\it Small Worlds} (Princeton University Press, Princeton, 1999).
\bibitem{pool_kochen} I. de Sola Pool and M. Kochen, Social Networks {\bf 1}, 1 (1978/79). 
\bibitem{was_faust} S. Wassermann and K. Faust, {\it Social Network Analysis} (Cambridge University Press, Cambridge, 1994). 
\bibitem{new_coll} M. E. J. Newman, Phys. Rev. E {\bf 64}, 016131 (2001). 
\bibitem{vazvesp} A. Vazquez, R. Pastor-Satorras, and A. Vespigniani, {Phys. Rev. E} {\bf 65}, 066130 (2002). 
\bibitem{rav_barab} E. Ravasz and A.-L. Barab\'asi, Phys. Rev. E {\bf 67}, 026112 (2003).
\bibitem{bollob} B. Bollob\'as, O. Riordan, J. Spencer, and G. Tusn\'ady, Rand. Struct. Alg. {\bf 18}, 279 (2001). 
\bibitem{jin} E. M. Jin, M. Girvan, and M. E. J. Newman, Phys. Rev. E 64, 046132 (2001).
\bibitem{vazquez} A. Vazquez, Europhys. Lett. {\bf 54}, 430 (2001).
\bibitem{hk} P. Holme and B. J. Kim, Phys. Rev. E. {\bf 65}, 026107 (2002). 
\bibitem{deb} J. Davidsen, H. Ebel, and S. Bornholdt, Phys. Rev. Lett. {\bf 88}, 128701 (2002). 
\bibitem{new_timedev} M. E. J. Newman, Phys. Rev. E {\bf 64}, 025102 (2001). 
\end{thebibliography}
\end{document}